\documentclass[12pt]{iopart}
\newcommand{\gguide}{{\it Preparing graphics for IOP journals}}
\usepackage{iopams}
\usepackage{graphicx}
\usepackage{amssymb}
\usepackage{amscd}
\usepackage{mathrsfs}
\usepackage{latexsym}
\usepackage{array}
\usepackage{amsthm}
\usepackage{subfigure}
\usepackage{multirow}

\usepackage{tikz}
\usetikzlibrary{matrix}

\usepackage[labelsep=newline,skip=2pt]{caption}

\theoremstyle{definition}

\newcommand{\set}[1]{\ensuremath{\{#1\}}}
\newcommand{\tc}{\ensuremath{\ |\ }}

 \newtheoremstyle{ax}
     {\topsep}
     {\topsep}
     {\itshape}
     {}
     {\bfseries}
     {.}
     {\topsep}
     {\thmname{#1}\thmnumber{#2}\thmnote{ (#3)}}
\theoremstyle{ax}

\newtheorem{esr}{ESR}[]
\newtheorem{qm}{QM}[]
\newtheorem{tm}{FM}[]
\newtheorem{p}{P}[]

\newtheoremstyle{am}
     {\topsep}
     {\topsep}
     {\itshape}
     {}
     {\bfseries}
     {.}
     {\topsep}
     {\thmname{#1}\thmnote{ (#3)}}
\theoremstyle{am}
\newtheorem{axio}{AX}[]
\newtheorem{ac}{AC}[]

\newtheorem{bm}{BM}[]
\newtheorem{im}{IM}[]

\begin{document}

\title[Finite Local Models for the GHZ Experiment]{Finite Local Models for the GHZ Experiment}

\author{C. Garola$^{1}$, M. Persano$^1$, J. Pykacz$^2$, S. Sozzo$^1$ $^3$}

\address{$^1$ Department of Mathematics and Physics, University of Salento \\
Via Arnesano, 73100 Lecce, Italy \\
INFN, Section of Lecce \\
Via Arnesano, 73100 Lecce, Italy}

\vspace*{-0.2cm}
\ead{Garola@le.infn.it, Persano@le.infn.it, Sozzo@le.infn.it}

\address{$^2$ Institute of Mathematics, University of Gdansk, Gdansk, Poland}

\vspace*{-0.2cm}
\eads{\mailto{pykacz@math.ug.edu.pl}}

\address{$^3$ Center Leo Apostel (CLEA), Free University of Brussels (VUB) \\
Krijgskundestraat 33, 1160 Brussels, Belgium}

\vspace*{-0.2cm}
\eads{\mailto{ssozzo@vub.ac.be}}

\begin{abstract}
Some years ago Szab\'o and Fine proposed a {\it local} hidden variable theory for the GHZ experiment based on the assumption that ``the detection efficiency is not (only) the effect of random errors in the detector equipment, but it is a more fundamental phenomenon, the manifestation of a predetermined hidden property of the particles''. Szab\'o and Fine, however, did not provide a general approach to quantum phenomena which avoids nonlocality. Such an approach, based on the same assumption, was instead recently supplied by some of us and called {\it extended semantic realism} ({\it ESR}) model. We show here that one can extract from the ESR model several local finite models referring to the specific physical situation considered in the GHZ experiment, and that these models can be converted into the toy models for the GHZ experiment worked out by Szab\'o and Fine.
\end{abstract}


\section{Introduction} \label{intro}
The \emph{Greenberger--Horne--Zeilinger} (\emph{GHZ}) argument \cite{ghz89} was devised more than twenty years ago to prove Bell's theorem (nonlocality of QM) without introducing inequalities. It is based on a gedankenexperiment which can be transformed into a real experiment (\emph{GHZ experiment}) \cite{ghsz90}.

Some years ago Szab{\'o} and Fine proved that a \emph{local} hidden variables theory can be supplied for the GHZ experiment based on the assumption that ``the detection efficiency is not (only) the effect of random errors in the detection equipment, but it is a more fundamental phenomenon, the manifestation of a predetermined hidden property of the particles'' \cite{sf02}. Their proof is actually based also on the further assumption that the probability measure $p$ to be defined on the space $\Lambda$ of the hidden variables of their theory must satisfy some constraints following from QM, which is equivalent to modifying the standard interpretation of quantum probabilities considering them as conditional on detection rather than absolute in the specific case of the GHZ experiment.

It is important to observe that, because of the above assumptions, the Szab{\'o} and Fine proof does not show that the GHZ argument is wrong. Rather, it shows that the GHZ argument depends on adopting the standard interpretation of quantum probabilities as absolute (in particular, when assuming perfect correlation, i.e., correlations with probability 1), while it fails to be true when this interpretation is modified in the sense proposed by Szab{\'o} and Fine. 

Szab{\'o} and Fine, however did not provide a general theory of quantum phenomena which avoids nonlocality. A theory of this kind was instead recently supplied by some of us and called \emph{extended semantic realism} ({\it ESR}) \emph{model} \cite{ga03,gp04,gs09,gs08,sg08,gs10,gs10b,gs10c,gs12}. The ESR model consists of a microscopic and a macroscopic part. The former can be considered as a hidden variables theory, which justifies and supports the assumptions stated in the latter providing a set--theoretical description of the microscopic world. The latter constitutes the part of the theory that is interpreted on the physical domain, and introduces a general mathematical apparatus which embodies the standard mathematical formalism of quantum mechanics (QM), but reinterprets quantum probabilities as conditional on detection rather than absolute, thus circumventing known ``no--go'' theorems \cite{b64,b66,ks67}. If one then compares the ESR model with the Szab{\'o} and Fine theory for the GHZ experiment, one immediately realizes that there is a correspondence between the assumptions of the two approaches. Indeed, the ESR model introduces in every measurement a {\it detection probability}, which depends on the values that the hidden variables ({\it microscopic states}) take on the individual example ({\it physical object}) of the physical system that is considered and not on flaws or lack of efficiency of the measuring apparatuses (hence the basic theoretical entities of the ESR model are {\it generalized observables} obtained by adding a {\it no--registration outcome}, which provides nontrivial information on the physical object that is measured, to the set of possible values of each standard observable of QM). This assumption restates and generalizes the first assumption of Szab{\'o} and Fine mentioned above, which follows an idea by Fine \cite{f82a,f82b,f89}, going back, according to Fine himself, to Einstein \cite{f94}. Furthermore,  we have seen that quantum probabilities are reinterpreted as conditional on detection in the ESR model, which restates and generalizes the second assumption of Szab{\'o} and Fine mentioned above. 

Based on the above remarks, we intend to show in this paper that several local finite models for the GHZ experiment can be extracted from the ESR model and immediately converted into the ``toy'' models supplied by Szab\'o and Fine for the same experiment, thus proving the consistency of our general proposal with some previous significant but very specific results in the literature. Hence we resume in Sec. \ref{esr} the notions in the ESR model that are required to reach this goal, leaving apart the mathematical formalism that is inessential to our aims. We then discuss in Sec. \ref{family} some properties that can be introduced to single out a family of local finite models from our general ESR model in the specific case of the GHZ experiment. Finally, we provide in Sec. \ref{toy} three examples of models in this family, showing that they can be converted into the Szab{\'o} and Fine toy models, as desired.

\section{The essentials of the ESR model}\label{esr}
According to the ESR model, 
a physical system $\Omega$ is operationally defined by a pair $(\Pi, {\mathscr R})$, with $\Pi$ a set of \emph{preparing devices} and ${\mathscr R}$ a set of \emph{measuring apparatuses}. Every preparing device, when activated, prepares an \emph{individual example} of $\Omega$. Every measuring device, if activated after a preparing device, yields an \emph{outcome}, that we assume to be a real number.

In the theoretical description a physical system $\Omega$ is characterized by a set ${\mathscr U}$ of \emph{physical objects}, a set $\mathscr{E}$ of \emph{microscopic properties} and a set ${\mathscr S}_{\mu} \subset {\mathscr{P}({\mathscr E})}$ (the power set of ${\mathscr E}$) of \emph{microscopic states} at a microscopic level, and by a set $\mathscr S$ of (\emph{macroscopic}) \emph{states} and a set ${\mathscr O}_{0}$ of \emph{generalized observables} at a macroscopic level.

Physical objects are operationally interpreted as individual examples of $\Omega$, while microscopic properties and states are purely theoretical entities (the hidden variables of the model). Every physical object $\alpha \in {\mathscr U}$ is associated with a microscopic state, \emph{i.e.}, a set $S_{\mu} \in {\mathscr S}_{\mu}$ of microscopic properties (the microscopic properties \emph{possessed} by $\alpha$). We briefly say that ``$\alpha$ is in the state $S_\mu$'', and denote by $extS_\mu$ the set of all physical objects in the state $S_\mu$ (hence $extS_\mu\subset\mathscr{U}$).
 
States are physically defined as classes of probabilistically equivalent preparing devices, following standard procedures in the foundations of QM \cite{bc81}. Every device $\pi \in S \in {\mathscr S}$, when constructed and activated, prepares an individual example of $\Omega$, hence a physical object $\alpha$. We briefly say that ``$\alpha$ is (prepared) in the state $S$'' and denote by $extS$ the set of all physical objects in the state $S$ (hence $extS\subset\mathscr{U}$). Analogously, generalized observables\footnote{One could obtain a more general theory by introducing \emph{unsharp} generalized observables at this stage. We do not consider this generalization of the ESR model in the present paper.}
 are physically defined as classes of probabilistically equivalent \emph{measuring apparatuses}. Every $A_0 \in {\mathscr O}_{0}$ is obtained by considering an observable $A$ in the set ${\mathscr O}$ of all observables of QM and adding a \emph{no--registration outcome} $a_0\in\Re$ to the set $\Xi$ of all possible values of $A$ on the real line $\Re$, so that the set of all possible values of $A_0$ is $\Xi_0=\{ a_0 \} \cup \Xi$.\footnote{We assume that $\Re \setminus \Xi$ is non--void, which is not restrictive. Indeed, if $\Xi=\Re$, one can choose a bijective Borel function $f: \Re \rightarrow \Xi'$ such that $\Xi' \subset \Re$, and replace $A$ with $f(A)$. \label{borel_sets}}
The set ${\mathscr F}_{0}$ of all (\emph{macroscopic}) \emph{properties} of $\Omega$ is then defined as follows,
\begin{equation}
{\mathscr F}_{0} \ =  \ \{ (A_0, X) \ | \  A_0 \in {\mathscr O}_{0}, \ X \in \mathbb{B}(\Re) \},
\end{equation}
where $\mathbb{B}(\Re)$ is the $\sigma$--algebra of all Borel subsets of $\Re$. Hence the subset 
\begin{equation}
{\mathscr F} \ =  \ \{ (A_0, X) \ | \  A_0 \in {\mathscr O}_{0}, \ X \in \mathbb{B}(\Re), \ a_0 \notin X \} \subset {\mathscr F}_{0},
\end{equation}
is in one--to--one correspondence with the set
\begin{equation}
{\mathscr G}=\{ (A, X) \ | \  A \in {\mathscr O}, \ X \in \mathbb{B}(\Re) \}
\end{equation}
of all quantum properties associated with observables of QM through the bijective mapping\footnote{We have usually identified the sets $\mathscr{F}$ and $\mathscr{G}$ in our earlier presentation of the ESR model. For the sake of clearness, we avoid such an identification in this section, and re--establish it in Sec. \ref{family}.}
\begin{equation}
q: (A_0,X)\in {\mathscr F} \mapsto (A,X) \in {\mathscr G},
\end{equation}
and we mainly deal with this subset in the following.

A measurement of a property $F=(A_0, X)$ on a physical object $\alpha$ in the state $S$ is then described as a \emph{registration} performed by means of a \emph{dichotomic registering device} whose outcomes are denoted by \emph{yes} and \emph{no}. The measurement yields outcome yes/no (equivalently, $\alpha$ \emph{displays}/\emph{does not display} $F$), if and only if the value of $A_0$ belongs/does not belong to $X$.

The connection between the microscopic and the macroscopic part of the ESR model is established by introducing the following  \emph{bijective mapping} (BM) and \emph{idealized measurements} (IM) assumptions.

\begin{bm}
A bijective mapping $\varphi: {\mathscr E} \longrightarrow {\mathscr F}$ exists.
\end{bm}

\begin{im}
If a physical object $\alpha$ is in the microscopic state $S_{\mu}$ and an \emph{idealized measurement} of a property $F=\varphi (f)\in\mathscr{F}$ is performed on $\alpha$, then $S_{\mu}$ determines a probability $p_{S_{\mu}}^{d}(F)$ that $\alpha$ be detected, and $\alpha$ displays $F$ if it is detected and $f \in S_{\mu}$, $\alpha$ does not display $F$ if it is not detected or $f \notin S_{\mu}$. 
\end{im}
\noindent For the sake of simplicity, we will consider only idealized measurements (briefly called \emph{measurements} from now on) in the following.

Assumption BM allows us to associate a microscopic state $S_{\mu}$ with the set $\{F=\varphi(f) \ | \ f \in S_{\mu}  \}$ or, equivalently, with the set $\{G=q(\varphi(f))\ | \ f \in S_{\mu} \}$. We call \emph{characterization of $S_{\mu}$} each of these sets of properties.

The ESR model is \emph{deterministic} if $p_{S_{\mu}}^{d}(F) \in \{0,1 \}$, \emph{probabilistic} otherwise. In the former case it is necessarily noncontextual: hence all properties are objective, because the outcome of the measurement of a property on a physical object $\alpha$ depends only on the microscopic properties possessed by $\alpha$ and not on the measurement context. In the latter case one can recover noncontextuality by adding further hidden variables which make $p_{S_{\mu}}^{d}(F)$ epistemic to the microscopic properties \cite{gs08}.\footnote{The idealized measurements introduced in the ESR model correspond to the \emph{ideal measurements} of QM, and noncontextuality refers to idealized measurements only (local contextuality can indeed occur in actual measurements, \emph{e.g.}, if a threshold exists \cite{a2009}).}
  
By using the connection between the microscopic and the macroscopic part of the ESR model one can show \cite{gs08} that, whenever the property $F=(A_0, X) \in \mathscr F$ is measured on a physical object $\alpha$ in the state $S$, the overall probability $p_{S}^{t}(F)$ that $\alpha$ display $F$ is given by
\begin{equation} \label{formuladipartenza}
p_{S}^{t}(F)= p_{S}^{d}(F)p_{S}(F).
\end{equation}
The symbol $p_{S}^{d}(F)$ in Eq.~(\ref{formuladipartenza}) denotes the probability that $\alpha$ be detected whenever it is in the state $S$ (\emph{detection probability}) and $F$ is measured. The value of $p_{S}^{d}(F)$ is not fixed for a given $A_0 \in {\mathscr O}_{0}$ because it may depend on $F$, hence on $X$. But the connection of microscopic with macroscopic properties via $\varphi$ implies that $p_{S}^{d}(F)$ can be expressed in terms of the microscopic features of the physical objects in the state $S$, hence it does not occur because of flaws or lack of efficiency of the apparatus measuring $F$.

The symbol $p_{S}(F)$ in Eq.~(\ref{formuladipartenza}) denotes the conditional probability that $\alpha$ display $F$ when it is detected.

Eq.~(\ref{formuladipartenza}) introduces three basic probabilities. We have as yet no theory which allows us to predict the value of $p_{S}^{d}(F)$. But we can consider $p_{S}^{d}(F)$ as an unknown parameter to be determined empirically and introduce theoretical assumptions that connect the ESR model with QM enabling us to provide  mathematical representations of the physical entities introduced in the ESR model together with explicit expressions of $p_{S}^{t}(F)$ and $p_{S}(F)$.

Let us begin with $p_{S}(F)$. Then, the following statement expresses the fundamental assumption of the ESR model.

\begin{axio}
Let $S$ be a pure state and let $F \in {\mathscr F}$. Then the probability $p_{S}(F)$ can be evaluated by using the same rules that yield the probability of the property $G=q(F)\in\mathscr{G}$ in the state $S$ according to QM.
\end{axio}
 
Assumption AX allows one to recover the basic formalism of QM in the framework of the ESR model but modifies its standard interpretation. Indeed, according to QM, whenever an ensemble ${\mathscr E}_{S}$ of physical objects in a state $S$ is prepared and ideal measurements of the property $G=q(F)$ are performed, all physical objects in ${\mathscr E}_{S}$ are detected, hence the quantum rules yield the probability that a physical object $\alpha$ display $G$ if $\alpha$ is selected in ${\mathscr E}_{S}$ (\emph{absolute} probability). According to assumption AX, instead, if $S$ is pure, the quantum rules yield the probability that a physical object $\alpha$ display the property $F=q^{-1}(G)$ if idealized measurements of $F$ are performed and $\alpha$ is selected in the subset of all objects of ${\mathscr E}_{S}$ that are detected (\emph{conditional on detection} probability).

Because of the above reinterpretation of quantum probabilities the predictions of the ESR model are different from those of QM. 

From a physical point of view, however, one can only choose a device $\pi\in\Pi$ and then prepare $\alpha$ by means of $\pi$, so that $\alpha$ is in the macroscopic state $S\in \mathscr{S}$ characterized by $\pi$. Hence, if we denote by $p(S_\mu|S)$ the probability that a physical object $\alpha$ is in the microscopic state $S_\mu$ whenever it is in the state $S$, assigning $S$ selects a subset 
\begin{equation}
\mathscr{S}_{\mu|S}=\set{S_\mu\in\mathscr{S}_\mu\tc  p(S_\mu|S)\neq 0}\subset\mathscr{S}_\mu
\end{equation} 
of possible microscopic states of $\alpha$ but does not specify the actual microscopic state of $\alpha$. Nevertheless, the quantum laws that hold because of assumption AX whenever conditional on detection probabilities  are considered impose some restrictions on $\mathscr{S}_{\mu|S}$ via the bijective mappings $q$ and $\varphi$, as follows.

\begin{qm}
 Let \mbox{$F\in\mathscr{F}$} be such that a measurement of \mbox{$G=q(F)\in\mathscr{G}$} on a physical object $\alpha$ in the state $S$ yields outcome \emph{yes} (\emph{no}) with certainty according to QM. If a measurement of $F$ is performed on $\alpha$ and $\alpha$ is detected, then the microscopic state $S_\mu\in\mathscr{S}_{\mu|S}$ of $\alpha$ is such that $f=\varphi^{-1}(F)\in S_\mu$ ($f=\varphi^{-1}(F)\notin S_\mu$).
 
\end{qm}

\begin{qm}
\label{qm2}
Let us identify every $G\in\mathscr{G}$ with a dichotomic observable whose possible values are 1 and 0 and let us relabel as 1 and 0 the outcomes  \emph{yes} and \emph{no}, respectively, of any property $F\in\mathscr{F}_0$. Let us consider a family $\set{G_i}_{i\in I}$, with $I=\set{1,2,...,n}$, of compatible quantum properties, and let 
\begin{equation}
h(G_1,G_2,...,G_n)=0 
\end{equation}
be an (empirical) quantum law  which holds for every physical object $\alpha$ in the state $S$. Hence, whenever ideal measurements of $G_1,G_2,...,G_n$ are performed on a physical object $\alpha$ in the state $S$ obtaining the outcomes $v(G_1),v(G_2),...,$ $v(G_n)$ respectively, QM predicts that the equation 
\begin{equation}
h\big(v(G_1),v(G_2),...,v(G_n)\big)=0
\end{equation}
holds. From the point of view of the ESR model the ideal measurements of $G_1,G_2,...,G_n$ correspond to idealized measurements of the properties $F_1=q^{-1}(G_1),\,F_2=q^{-1}(G_2),\,...,\,F_n=q^{-1}(G_n)$, respectively. If $\alpha$ is detected in every measurement, the obtained outcomes are $v(F_1)=v(G_1),\,v(F_2)=v(G_2),\,...,\,v(F_n)=v(G_n)$, hence the equation 
\begin{equation}\label{hv}
h\big(v(F_1),v(F_2),...,v(F_n)\big)=0
\end{equation}
holds. Since assumption IM implies that, if $\alpha$ is detected, the microscopic property $f_i=\varphi^{-1}(F_i)$ ($i=1,2,...,n$) belongs to the microscopic state $S_\mu$  of $\alpha$ iff $v(F_i)=1$ (hence $f_i$ does not belong to $S_{\mu}$ iff $v(F_i)=0$), Eq.~\emph{(\ref{hv})} implies restrictions on the microscopic properties that can simultaneously belong to $S_\mu$ if $S_\mu\in\mathscr{S}_{\mu|S}$: hence, on $\mathscr{S}_{\mu|S}$.
 
\end{qm}

\section{A family of finite noncontextual models for the GHZ experiment}\label{family}
We consider in this section a physical system $\Omega$ made up of three spin--1/2 particles. A family of finite models for the GHZ experiment can then be extracted from the general ESR model for $\Omega$ by selecting finite sets of quantum observables. To this end, we consider three orthogonal directions $x,y,z$, use standard symbols to denote spin operators, and introduce the following sets of physical entities.

A reduced set $\tilde{{\mathscr O}}$ of quantum observables,
\begin{equation}
\tilde{{\mathscr O}}= \{ \sigma_{*}(n) \ |  \ *=x,y,z; \ n=1,2,3  \}
\end{equation}
(hence $\textrm{Card} \ \tilde{{\mathscr O}}=9$).

A reduced set $\tilde{{\mathscr G}}$ of quantum properties,
\begin{equation}
\tilde{{\mathscr G}}= \{ G_{*l}(n)=( \sigma_{*}(n), \{ l \} ) \ |  \ *=x,y,z; \ n=1,2,3; \ l=\pm 1  \}
\end{equation}
(hence $\textrm{Card} \ \tilde{{\mathscr G}}=18$). 

Following on the general rules of the ESR model, we then introduce a reduced set $\tilde{{\mathscr O}}_{0}= \{ \sigma_{*0}(n) \ |  \ *=x,y,z; \ n=1,2,3  \}$ of generalized observables, a reduced set $\tilde{{\mathscr F}}_{0}= \{ F_{*l}(n)=( \sigma_{*0}(n), \{ l \} ) \ |  \ *=x,y,z; \ n=1,2,3; \ l=\pm1,0  \}$ of (macroscopic) properties, a reduced subset $\tilde{{\mathscr F}}=q^{-1}(\tilde{{\mathscr G}})=\{ F_{*l}(n)=( \sigma_{*0}(n), \{ l \} ) \ |  \ *=x,y,z; \ n=1,2,3; \ l=+1,-1  \}$ of properties, and a reduced subset $\tilde{{\mathscr E}}={\varphi}^{-1}(\tilde{{\mathscr F}})$ of microscopic properties. However, we agree to simplify our models by identifying the sets $\tilde{{\mathscr E}}$ and $\tilde{{\mathscr F}}$ with $\tilde{{\mathscr G}}$ from now on, resting on the bijectivity of the mappings $q$ and $\varphi$.

The first assumption to be fulfilled by every model in our family limits the set of microscopic states that may occur in each model, as follows.
\begin{tm}\label{fm1}
The set of microscopic states in each model is given by
\begin{eqnarray}
\tilde{{\mathscr S}}_{\mu}= \Big \{ S_{\mu}=\{ (\sigma_{x}(1),\{i_1\}), (\sigma_{y}(1),\{j_1\}), (\sigma_{z}(1),\{k_1\});  (\sigma_{x}(2),\{i_2\}), \nonumber \\
 (\sigma_{y}(2),\{j_2\}), (\sigma_{z}(2),\{k_2\}); (\sigma_{x}(3),\{i_3\}), (\sigma_{y}(3),\{j_3\}), (\sigma_{z}(3),\{k_3\}) \} \nonumber \\
 \tc i_1,j_1,k_1,i_2,j_2,k_2,i_3,j_3,k_3=\pm 1 \Big \}.
\end{eqnarray}
\end{tm}

\noindent
{\it Justification.} The quantum properties $G_{*+1}(n)=( \sigma_{*}(n),\{+1\})$ and $G_{*-1}(n)=( \sigma_{*}(n),\{-1\})$, if considered as dichotomic observables (see QM2) are compatible and such that $G_{*+1}(n)+G_{*-1}(n)=I$. Let (ideal) measurements of $G_{*+1}(n)$ and $G_{*-1}(n)$ be performed on the n--th particle of a physical object $\alpha$. From the point of view of the ESR model, $\alpha$ is in a microscopic state $S_{\mu}$, and if the n--th particle is detected in both measurements, then one and only one of the quantum properties $G_{*+1}(n)$ and $G_{*-1}(n)$ belongs to the characterization of $S_{\mu}$ by means of quantum properties because of QM\ref{qm2}. Therefore, if $S_{\mu}$ is such that its characterization by means of quantum properties contains both $G_{*+1}(n)$ and $G_{*-1}(n)$, or none of them, then necessarily the n--th particle cannot be detected in one, or both, the measurements. It is then reasonable to agree that such microscopic states cannot occur in our model. We are thus left with those microscopic states only that are characterized by a triple of quantum properties for each particle.

\vspace{.2cm}
\noindent
Because of assumption FM\ref{fm1}, every $S_{\mu} \in \tilde{{\mathscr S}}_{\mu}$ can be characterized by a 9--tuple $( i_1,j_1,k_1;i_2,j_2,k_2;i_3,j_3,k_3)$.
For the sake of brevity, we will usually identify the microscopic state $S_{\mu}$ with this characterization in the following.

We can now state the second assumption to be fulfilled by every model in our family.
\begin{tm}\label{fm2}
For every $S_{\mu} \in \tilde{{\mathscr S}}_{\mu}$ and $G_{*l}(n) \in \tilde{\mathscr{G}}$, there exists a random variable
\begin{equation}
\lambda(S_{\mu}, G_{*l}(n)): ext S_{\mu} \longrightarrow \{ D, U \} 
\end{equation}
and the following equation holds
\begin{equation} \label{fm2assumption}
\lambda(S_{\mu}, G_{*+1}(n))= \lambda(S_{\mu}, G_{*-1}(n)).
\end{equation}
\end{tm}

\noindent
{\it Justification.} In the ESR model, every microscopic state $S_{\mu}$ determines a probability $p_{S_{\mu}}^{d}(F)$ that a physical object in the microscopic state $S_{\mu}$ is detected whenever a measurement of a macroscopic property $F$ is performed. Equivalently, for every pair $(S_{\mu},F)$ a random variable is defined on $extS_{\mu}$ which takes two values, $D$ ({\it detected}) and  $U$ ({\it undetected}). In our specific case all quantum observables are dichotomic, hence we can assume that the random variables associated with $G_{*+1}(n)$ and $G_{*-1}(n)$ are identical. 

\vspace{.2cm}
\noindent
Because of Eq. (\ref{fm2assumption}) we briefly write $\lambda(S_{\mu}, G_{*+1}(n))=\lambda_{*n}= \lambda(S_{\mu}, G_{*-1}(n))$, understanding the dependence on $S_{\mu}$. Hence every $S_{\mu} \in \tilde{{\mathscr S}}_{\mu}$ can be associated with a 9--tuple 
\begin{equation}
\Lambda(S_{\mu})=(\lambda_{x1}, \lambda_{y1}, \lambda_{z1}; \lambda_{x2}, \lambda_{y2}, \lambda_{z2}; \lambda_{x3}, \lambda_{y3}, \lambda_{z3})
\end{equation}
of random variables, called \emph{detection mappings}, such that 
\begin{equation}
\lambda_{*n}: extS_{\mu} \longrightarrow \{ D, U \}\quad (*=x,y,z; n=1,2,3).
\end{equation}
Moreover, because of assumption FM\ref{fm2}, we can consider measurements of the dichotomic quantum observable $\sigma_*(n)$ rather than measurements of the properties $G_{*+1}(n)$ and $G_{*-1}(n)$. Therefore, let a physical object $\alpha$ be in the microscopic state $S_{\mu}=(i_1,j_1,k_1;i_2,j_2,k_2;i_3,j_3,k_3)$, and let a measurement of the quantum observable $\sigma_{*}(n)$ be performed on $\alpha$. If $\lambda_{*n}(\alpha)=D$, then the n--th particle belonging to $\alpha$ is detected and the pointer of the measuring apparatus moves, yielding as outcome the value $v(\sigma_{*}(n))$ of $\sigma_{*}(n)$. If instead $\lambda_{*n}(\alpha)=U$, then the pointer remains in its initial position, that we label by 0. From the point of view of the ESR model we are actually measuring the generalized observable $\sigma_{*0}(n)$, which has three possible values (that is, $+1$, $0$ and $-1$), and get the outcome $v(\sigma_{*}(n))$ if $\lambda_{*n}(\alpha)=D$, the outcome $0$ if $\lambda_{*n}(\alpha)=U$. Therefore the outcomes of any possible set of measurements on $\alpha$ can be obtained by assigning a pair $(S_{\mu}, \lambda(\alpha))$, where
\begin{eqnarray}\lambda(\alpha)\!=\!(\mbox{\fontsize{10}{13}\selectfont$
\lambda_{x1}(\alpha), \lambda_{y1}(\alpha), \lambda_{z1}(\alpha);
 \lambda_{x2}(\alpha), \lambda_{y2}(\alpha), \lambda_{z2}(\alpha); \lambda_{x3}(\alpha), \lambda_{y3}(\alpha), \lambda_{z3}(\alpha)$})
\end{eqnarray}
is the {\it detection distribution} (briefly, {\it d--distribution}) of $\alpha$ (note that $\alpha,\alpha'\in ext S_\mu$ does not generally imply $\lambda(\alpha)=\lambda(\alpha')$).

The resulting set of outcomes can then be collected in a \emph{measurement specification} (briefly, \emph{m--specification}) associated with $\alpha$,
\begin{equation}
m(\alpha)=(r_1,s_1,t_1;r_2,s_2,t_2;r_3,s_3,t_3)
\end{equation}
where $r_1=i_1$ iff $\lambda_{x1}(\alpha)=D$, $r_1=0$ otherwise,  $s_1=j_1$ iff $\lambda_{y1}(\alpha)=D$, $s_1=0$ otherwise, etc. 

Let us come to (macroscopic) states. As we have anticipated in Sec.~\ref{intro}, we are interested in this paper in providing noncontextual (hence local) finite models which predict the results of a GHZ experiment in accordance with QM. Hence we consider in the following only one state $S \in \mathscr{S}$, the \emph{GHZ} state \cite{ghz89,ghsz90} represented by the vector 
\begin{equation}
|\psi\rangle=\frac{1}{\sqrt{2}}(|+1,+1,+1\rangle_{z}-|-1,-1,-1\rangle_{z})
\end{equation}
of the Hilbert space associated with $\Omega$ in QM. Whenever a physical object $\alpha$ is in the state $S$ (equivalently, $\alpha \in ext S$), joint measurements of $\sigma_{z}(1)$, $\sigma_{z}(2)$ and $\sigma_{z}(3)$ yield either the triple of outcomes $(+1,+1,+1)$ or the triple $(-1,-1,-1)$. This suggests introducing the following further assumption in our family of finite models.
\begin{tm}\label{fm3}
$\ \tilde{{\mathscr S}}_{\mu|S}=\{S_{\mu} \in \tilde{{\mathscr S}}_{\mu} \ | \ p(S_{\mu}|S) \ne 0 \}=$
\begin{equation}
=\{ (i_1,j_1,k;i_2,j_2,k;i_3,j_3,k) \ | \ i_1,j_1,i_2,j_2,i_3,j_3,k=\pm 1 \}
\end{equation}
(hence $N=\textrm{Card} \ \tilde{{\mathscr S}}_{\mu|S}=2^{7}=128$). Equivalently, $\tilde{{\mathscr S}}_{\mu|S}$ contains all microscopic states in $\tilde{{\mathscr S}}_{\mu}$ such that $k_1=k_2=k_3=k$ and only those.
\end{tm}

\noindent
{\it Justification.} Simplicity of the models.

\vspace{.2cm}
\noindent
We must now introduce probability within our models. To this end, let us note preliminarily that, for every physical object $\alpha \in S_{\mu}$, the d--distribution $\lambda(\alpha)$ must satisfy some restrictions imposed by the laws of QM, as we shall presently see, that is, must be physically possible. Then, $\Lambda(S_{\mu})$ assigns a probability $p(\lambda(\alpha))$ to each physically possible d--distribution (we stress that we have not assumed that the detection mappings are independent random variables: hence, generally, $p(\lambda(\alpha))$ is not a trivial product of independent factors, see Secs. \ref{3toy} and \ref{2toy}, though it may be such in special cases, see Sec. \ref{1toy}). Bearing in mind the symbols introduced in Sec. \ref{esr}, we therefore state the following assumption.
\begin{tm}\label{fm4}
(i) For every $S_{\mu}, S'_{\mu} \in \tilde{{\mathscr S}}_{\mu|S}$, $p(S_{\mu}|S)=p(S'_{\mu}|S)$.

(ii) Let $S_{\mu} \in \tilde{{\mathscr S}}_{\mu|S}$, $\alpha\in ext S_{\mu}$, $\lambda, \lambda' \in \Lambda(S_{\mu})$. Then, $p(\lambda(\alpha))=p(\lambda'(\alpha))$.
\end{tm}

\vspace{.2cm}
\noindent
{\it Justification.} Statement FM\ref{fm4}, (i) rests on the intuitive idea that, whenever a huge number of physical objects in the state $S$ are produced, they distribute uniformly in the microscopic states of $\tilde{{\mathscr S}}_{\mu|S}$. Statement FM\ref{fm4}, (ii) makes our family of finite models as simple as possible, postulating uniform distribution also on the set of all d--distributions that are compatible with the laws of QM when the state $S_{\mu}$ is given.

\vspace{.2cm}
\noindent
Assumption FM\ref{fm4} has some obvious consequences that will be widely used in Sec. \ref{toy}. Indeed, statement (i) in FM\ref{fm4} implies that, for every $S_\mu\in\tilde{\mathscr{S}}_{\mu|S}$, $p(S_\mu|S)=1/N=1/128$. Furthermore, statement (ii) in FM\ref{fm4} implies that, for every  $\alpha\in ext S_\mu$, $p(\lambda(\alpha))=1/{d(S_\mu)}$, where $d(S_\mu)$ is the number of d--distributions that are physically possible in the microscopic state $S_\mu$.

Assumptions FM\ref{fm1}--FM\ref{fm4} are fulfilled by a huge class of finite models. We select our family of finite models by accepting in it only the models satisfying the following {\it adequacy condition} (AC), which supplies them with a physical meaning.

\begin{ac}
For every $\alpha \in ext S$ the conditional on detection probability of obtaining a given set of outcomes when compatible measurements are performed on $\alpha$ must coincide with the probability predicted by QM.
\end{ac}

\noindent
{\it Justification.} Assumption AX in Sec.~\ref{esr} states that the conditional on detection probabilities introduced by the ESR model must coincide with the probabilities predicted by QM \cite{ga03,gp04,gs09,gs08,gs10}. The adequacy condition AC then transfers this general assumption to the finite models in our family.

\vspace{.2cm}
\noindent
Condition AC imposes several restrictions on the detection mappings. In particular, a straightforward quantum calculation shows that the conditional on detection probability of obtaining a given outcome ($+1$ or $-1$) when measuring an arbitrary observable on a particle of $\alpha$ must be $1/2$. The conditional on detection probability of obtaining a given combination of the outcomes $+1$ and $-1$ when measuring two arbitrary observables on two different particles of $\alpha$ must be $1/4$, but when both observables are spin components along the z-axis, in which case it must be $1/2$ if the outcomes have the same sign, $0$ if the outcomes have opposite signs. The restriction imposed on the conditional on detection probability of obtaining a given combination of the outcomes $+1$ and $-1$ when measuring three arbitrary observables on the three particles of $\alpha$ are less obvious. In particular, if three (two) observables are spin components along the z-axis, the foregoing probability must be $1/2$ ($1/4$) if all outcomes have the same sign, $0$ ($0$) if the signs of the outcomes are different. Instead, if only one of the three observables is a spin component along the z--axis, then the conditional on detection probability must be $1/8$. This value of the conditional on detection probability must also be predicted by our models whenever one performs a joint measurement of $\sigma_{x}(1)$, $\sigma_{x}(2)$ and $\sigma_{y}(3)$, or $\sigma_{x}(1)$, $\sigma_{y}(2)$ and $\sigma_{x}(3)$, or $\sigma_{y}(1)$, $\sigma_{x}(2)$ and $\sigma_{x}(3)$, or $\sigma_{y}(1)$, $\sigma_{y}(2)$ and $\sigma_{y}(3)$. But if one considers joint measurements of the following triples of compatible observables in QM
\begin{eqnarray}
M^{I}&=\{ \sigma_{x}(1), \sigma_{y}(2), \sigma_{y}(3) \}, \\
M^{II}&=\{ \sigma_{y}(1), \sigma_{x}(2), \sigma_{y}(3) \}, \\
M^{III}&=\{ \sigma_{y}(1), \sigma_{y}(2), \sigma_{x}(3) \}, \\
M^{IV}&=\{ \sigma_{x}(1), \sigma_{x}(2), \sigma_{x}(3) \}
\end{eqnarray}
(briefly, the \emph{measurements} $M^{I}, M^{II}, M^{III}$ and $M^{IV}$) one obtains that the probabilities of getting the outcomes $v(\sigma_x(1))=i_1$, $v(\sigma_y(2))=j_2$ and $v(\sigma_y(3))=j_3$ in $M^{I}$, $v(\sigma_y(1))=j_1$, $v(\sigma_x(2))=i_2$ and $v(\sigma_y(3))=j_3$ in $M^{II}$,  $v(\sigma_y(1))=j_1$, $v(\sigma_y(2))=j_2$ and $v(\sigma_x(3))=i_3$ in $M^{III}$, and $v(\sigma_x(1))=i_1$, $v(\sigma_x(2))=i_2$ and $v(\sigma_x(3))=i_3$ in $M^{IV}$ are given by
\begin{eqnarray}
p_{i_1,j_2,j_3}^{\psi}&=\raisebox{-0.3ex}{\fontsize{17.28}{20}\selectfont$\frac{1}{8}$}(1+i_1j_2j_3), \label{I}\\
p_{j_1,i_2,j_3}^{\psi}&=\raisebox{-0.3ex}{\fontsize{17.28}{20}\selectfont$\frac{1}{8}$}(1+j_1i_2j_3), \label{II}\\
p_{j_1,j_2,i_3}^{\psi}&=\raisebox{-0.3ex}{\fontsize{17.28}{20}\selectfont$\frac{1}{8}$}(1+j_1j_2i_3), \label{III}\\
p_{i_1,i_2,i_3}^{\psi}&=\raisebox{-0.3ex}{\fontsize{17.28}{20}\selectfont$\frac{1}{8}$}(1-i_1i_2i_3), \label{IV}
\end{eqnarray}
respectively. Eqs. (\ref{I})--(\ref{IV}) imply that the outcomes that one obtains when performing $M^{I}$, or $M^{II}$, or $M^{III}$, or $M^{IV}$ must fulfill the following equations
\begin{eqnarray}
i_1j_2j_3=+1, \label{vI}\\
j_1i_2j_3=+1, \label{vII}\\
j_1j_2i_3=+1, \label{vIII}\\
i_1i_2i_3\hspace{0.5mm}=-1, \label{vIV}
\end{eqnarray}
respectively. It has been widely commented in the literature on the fact that Eqs. (\ref{vI})--(\ref{vIV}) cannot be fulfilled simultaneously. In the orthodox view this impossibility shows that one cannot assume that the values of the observbles that occur in these equations are predetermined. In different words, one can rest on Eqs. (\ref{vI})--(\ref{vIV}) to get a straightforward proof of the Bell theorem (contextuality at a distance, or {\it nonlocality}). From the point of view of the ESR model, instead, Eqs. (\ref{vI})--(\ref{vIV}) entail some restrictions on the detection mappings. To be precise, they require that the following \emph{detection mappings} (DM) condition, which follows from condition AC, be fulfilled.

\vspace{.2cm}
\noindent
\textbf{DM.} \emph{Let a measurement $M^{I}$, or $M^{II}$, or $M^{III}$, or $M^{IV}$ be performed on a physical object $\alpha \in ext S_{\mu}$, with $S_{\mu} \in \tilde{{\mathscr S}}_{\mu|S}$. Then the detection mappings must be such that at least one of the particles of $\alpha$ remains undetected whenever the values of the observables that characterize $S_{\mu}$ do not fulfill the equation corresponding to the measurement that is performed.}

\vspace{.2cm}
\noindent
Condition DM (or, more generally, condition AC) is not sufficient to determine all detection mappings. According to the ESR model these mappings are actually determined by the physical system that is considered, but the ESR model does not provide a general theory for them. Hence we can supply a family of finite models which are distinguished by different choices of the detection mappings. Every model in the family is then characterized by the set of all m--specifications associated with physical objects in $ext S$, which must be such that condition AC (hence Eqs.  (\ref{vI})--(\ref{vIV})) is satisfied.

To close this section we stress that assumption FM\ref{fm2} implies that $\lambda_{*n}(\alpha)$ does not depend on the measurements that are performed on $\alpha$. Hence the detection mappings in our models must satisfy condition DM without depending on the choice of the measurement ($M^{I}$, or $M^{II}$, or $M^{III}$, or $M^{IV}$). This consequence of FM2 is physically important. Indeed, dependence on the choice of the measurement would imply a new form of nonlocality, because far away measurements on one of the particles in $\alpha$ would influence detection in the measurements on other particles.

\section{Examples of finite models for the GHZ experiment}\label{toy}
As we have anticipated in Sec. \ref{intro}, we intend to exhibit in this section some finite models for the GHZ experiment that belong to the family introduced in Sec. \ref{family}, and to show that these models can be easily converted into the toy models proposed by  Szab\'{o} and Fine for the same experiment \cite{sf02}. To this end, it is expedient to refer to a partition of $\tilde{{\mathscr S}}_{\mu|S}$ induced by the measurements $M^{I}, M^{II}, M^{III}$ and $M^{IV}$ introduced in Sec. \ref{family}, as follows.

The measurements $M^{I}, M^{II}, M^{III}$ and $M^{IV}$ can be associated with the subsets 
\begin{equation}
{\hspace*{-2cm}}\tilde{{\mathscr S}}_{\mu|S}^{I}=\{ S_{\mu}=(i_1,j_1,k;i_2,j_2,k;i_3,j_3,k) \ | \ i_1,j_1,i_2,j_2,i_3,j_3,k=\pm 1,  i_1j_2j_3=+1 \} \\
\end{equation}
\begin{equation}
{\hspace*{-2cm}}\tilde{{\mathscr S}}_{\mu|S}^{II}=\{ S_{\mu}=(i_1,j_1,k;i_2,j_2,k;i_3,j_3,k) \ | \ i_1,j_1,i_2,j_2,i_3,j_3,k=\pm 1,  j_1i_2j_3=+1 \} \\
\end{equation}
\begin{equation}
{\hspace*{-2cm}}\tilde{{\mathscr S}}_{\mu|S}^{III}=\{ S_{\mu}=(i_1,j_1,k;i_2,j_2,k;i_3,j_3,k) \ | \ i_1,j_1,i_2,j_2,i_3,j_3,k=\pm 1,  j_1j_2i_3=+1 \} \\
\end{equation}
\begin{equation}
{\hspace*{-2cm}}\tilde{{\mathscr S}}_{\mu|S}^{IV}=\{ S_{\mu}=(i_1,j_1,k;i_2,j_2,k;i_3,j_3,k) \ | \ i_1,j_1,i_2,j_2,i_3,j_3,k=\pm 1,  i_1i_2i_3=-1 \}
\end{equation}
respectively (note that $\textrm{Card} \ \tilde{{\mathscr S}}_{\mu|S}^{I}= \textrm{Card} \ \tilde{{\mathscr S}}_{\mu|S}^{II}= \textrm{Card} \ \tilde{{\mathscr S}}_{\mu|S}^{III}= \textrm{Card} \ \tilde{{\mathscr S}}_{\mu|S}^{IV}=2^{6}=64$). The following statements then hold.
\begin{p}
{\hspace*{-2cm}}
\begin{equation} \label{cap}
\tilde{{\mathscr S}}_{\mu|S}^{I}\cap  \tilde{{\mathscr S}}_{\mu|S}^{II} \cap \tilde{{\mathscr S}}_{\mu|S}^{III} \cap \tilde{{\mathscr S}}_{\mu|S}^{IV}=\emptyset.
\end{equation}
Proof. \emph{Eqs.  (\ref{vI})--(\ref{vIV}) cannot be fulfilled simultaneously}.
\end{p}
\begin{p}
{\hspace*{-2cm}}
\begin{equation} \label{cup}
\tilde{{\mathscr S}}_{\mu|S}^{I}\cup  \tilde{{\mathscr S}}_{\mu|S}^{II} \cup \tilde{{\mathscr S}}_{\mu|S}^{III} \cup \tilde{{\mathscr S}}_{\mu|S}^{IV}=\tilde{{\mathscr S}}_{\mu|S}.
\end{equation}
Proof. \emph{The sets $\tilde{{\mathscr S}}_{\mu|S} \setminus \tilde{{\mathscr S}}_{\mu|S}^{I}$, $\tilde{{\mathscr S}}_{\mu|S} \setminus \tilde{{\mathscr S}}_{\mu|S}^{II}$, $\tilde{{\mathscr S}}_{\mu|S} \setminus \tilde{{\mathscr S}}_{\mu|S}^{III}$ and $\tilde{{\mathscr S}}_{\mu|S} \setminus \tilde{{\mathscr S}}_{\mu|S}^{IV}$ are characterized by equations analogous to Eqs.  (\ref{vI})--(\ref{vIV}), respectively, in which the signs in the right members are changed. Also these equations cannot be fulfilled simultaneously, which implies
\begin{equation}
(\tilde{{\mathscr S}}_{\mu|S} \setminus \tilde{{\mathscr S}}_{\mu|S}^{I}) \cap (\tilde{{\mathscr S}}_{\mu|S} \setminus \tilde{{\mathscr S}}_{\mu|S}^{II}) \cap (\tilde{{\mathscr S}}_{\mu|S} \setminus \tilde{{\mathscr S}}_{\mu|S}^{III}) \cap (\tilde{{\mathscr S}}_{\mu|S} \setminus \tilde{{\mathscr S}}_{\mu|S}^{IV}) = \emptyset
\end{equation}
or, equivalently, Eq.~(\ref{cup}).}
\end{p}
\begin{p}\label{p3}
Let us put $\tilde{{\mathscr S}}_{\mu|S}^{I0}=\tilde{{\mathscr S}}_{\mu|S} \setminus (\tilde{{\mathscr S}}_{\mu|S}^{II} \cup \tilde{{\mathscr S}}_{\mu|S}^{III} \cup \tilde{{\mathscr S}}_{\mu|S}^{IV})$, $\tilde{{\mathscr S}}_{\mu|S}^{II0}=\tilde{{\mathscr S}}_{\mu|S} \setminus (\tilde{{\mathscr S}}_{\mu|S}^{I} \cup \tilde{{\mathscr S}}_{\mu|S}^{III} \cup \tilde{{\mathscr S}}_{\mu|S}^{IV})$, $\tilde{{\mathscr S}}_{\mu|S}^{III0}=\tilde{{\mathscr S}}_{\mu|S} \setminus (\tilde{{\mathscr S}}_{\mu|S}^{I} \cup \tilde{{\mathscr S}}_{\mu|S}^{II} \cup \tilde{{\mathscr S}}_{\mu|S}^{IV})$, and $\tilde{{\mathscr S}}_{\mu|S}^{IV0}=\tilde{{\mathscr S}}_{\mu|S} \setminus (\tilde{{\mathscr S}}_{\mu|S}^{I} \cup \tilde{{\mathscr S}}_{\mu|S}^{II} \cup \tilde{{\mathscr S}}_{\mu|S}^{III})$. Then, the family
\begin{eqnarray}
{\mathscr P}= \{ \tilde{{\mathscr S}}_{\mu|S}^{I0}, \tilde{{\mathscr S}}_{\mu|S}^{II0}, \tilde{{\mathscr S}}_{\mu|S}^{III0}, \tilde{{\mathscr S}}_{\mu|S}^{IV0}, \tilde{{\mathscr S}}_{\mu|S}^{I} \cap \tilde{{\mathscr S}}_{\mu|S}^{II} \cap \tilde{{\mathscr S}}_{\mu|S}^{III}, \tilde{{\mathscr S}}_{\mu|S}^{I}\cap  \tilde{{\mathscr S}}_{\mu|S}^{II} \cap \tilde{{\mathscr S}}_{\mu|S}^{IV}, \nonumber \\
 \tilde{{\mathscr S}}_{\mu|S}^{I}\cap  \tilde{{\mathscr S}}_{\mu|S}^{III} \cap \tilde{{\mathscr S}}_{\mu|S}^{IV}, \tilde{{\mathscr S}}_{\mu|S}^{II}\cap  \tilde{{\mathscr S}}_{\mu|S}^{III} \cap \tilde{{\mathscr S}}_{\mu|S}^{IV} \}
\end{eqnarray}
is a partition of $\tilde{{\mathscr S}}_{\mu|S}$ (note that each set in ${\mathscr P}$ has cardinality $2^{4}=16$). \\
Proof. \emph{Straightforward from Eqs. (\ref{cap}) and (\ref{cup}).}
\end{p}

\subsection{$\mathscr{M}(3)$: a finite model with three detection failures}\label{1toy}
Table 1 displays a finite model, that we agree to call $\mathscr{M}(3)$, which has the following properties. 

\begin{figure} 
\begin{center}
\includegraphics[width=0.6\textwidth]{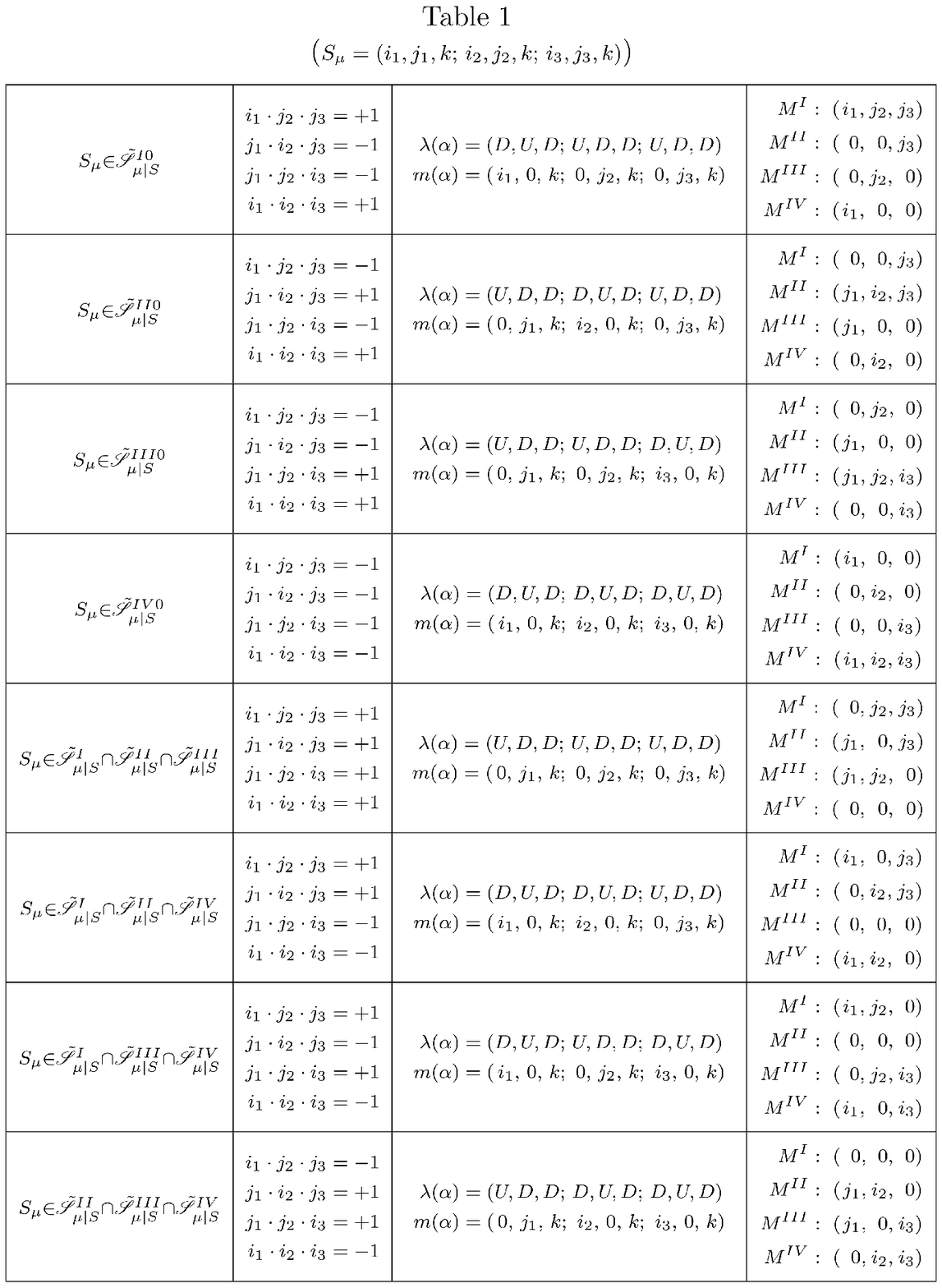}
\end{center}
\end{figure}

\vspace*{0.2cm}
\noindent
(i) Let $S_\mu,S'_\mu\in\tilde{\mathscr{S}}_{\mu|S}$, $\alpha \in ext S_\mu$, $\alpha' \in ext {S'_\mu}$. Then $\lambda(\alpha)=\lambda(\alpha')$ 
whenever $S_\mu$ and ${S'_\mu}$ belong to the same element of the partition in P3.

\vspace{0.2cm}
\noindent
(ii) For every physical object $\alpha$, three detection mappings take value $U$.

\vspace{0.2cm}
\noindent
It follows from (i) that, for every $S_\mu\in\tilde{\mathscr{S}}_{\mu|S}$ and $\alpha,\alpha'\in ext S_\mu$, $\lambda(\alpha)=\lambda(\alpha')$, which implies that there is only one possible d--distribution that can be associated with $S_\mu$, whose probability is then $1$ because of FM\ref{fm4}, (ii). Hence $S_{\mu}$ determines whether a physical object $\alpha \in ext S_\mu$ is detected or not when a measurement of $\sigma_*(n)$ is performed. $\mathscr{M}(3)$ is therefore \emph{deterministic} according to the terminology introduced in the ESR model (Sec.~\ref{esr}). Furthermore, microscopic states $S_\mu$ and $S'_\mu$ in different elements of the partition in P3 are associated with different d--distributions. 

Coming to m--specifications, direct inspection shows that different microscopic states in the same element of the partition in P3 may be associated with the same m--specification. Indeed, the 16 microscopic states in $\tilde{\mathscr{S}}_{\mu|S}^{I0}$ are associated with 8 different m-specifications (to be precise, $(+1,0,k;\,0,+1,k;$ $0,+1,k)$, $(+1,0,k;\,0,-1,k;\,0,-1,k)$, $(-1,0,k;\,0,+1,k;\,0,-1,k)$ and $(-1,0,$ $k;\,0,-1,k;\,0,+1,k)$, with $k=\pm 1$). Similarly, the 16 microscopic states in $\tilde{\mathscr{S}}_{\mu|S}^{II0}$, or $\tilde{\mathscr{S}}_{\mu|S}^{III0}$, or $\tilde{\mathscr{S}}_{\mu|S}^{IV0}$ are associated with 8 different m-specifications. The 16 microscopic states in $\tilde{\mathscr{S}}_{\mu|S}^{I} \cap \tilde{\mathscr{S}}_{\mu|S}^{II} \cap \tilde{\mathscr{S}}_{\mu|S}^{III}$ are associated instead with 16 different m-specifications (to be precise, $(0,+1,k;\,0,+1,k;\,0,+1,k)$, $(0,+1,k;\,0,-1,k;\,0,-1,k)$, $\,(0,+1,k;\,0,+1,k;\,0,-1,k)$, $\,(0,+1,k;\,0,-1,k;$ $0,+1,k)$, $\,(0,-1,k;\,0,-1,k;\,0,-1,k)$, $\,(0,-1,k;\,0,+1,k;\,0,+1,k)$, $\,(0,-1,k;$ $0,-1,k;\,0,+1,k)\,$ and $\,(0,-1,k;\,0,+1,k;\,0,-1,k)$, with $k=\pm 1$). Similarly, the 16 microscopic states in  $\tilde{\mathscr{S}}_{\mu|S}^{I} \cap \tilde{\mathscr{S}}_{\mu|S}^{II} \cap \tilde{\mathscr{S}}_{\mu|S}^{IV}$, or $\tilde{\mathscr{S}}_{\mu|S}^{I} \cap \tilde{\mathscr{S}}_{\mu|S}^{III} \cap \tilde{\mathscr{S}}_{\mu|S}^{IV}$, or $\tilde{\mathscr{S}}_{\mu|S}^{II} \cap \tilde{\mathscr{S}}_{\mu|S}^{III} \cap \tilde{\mathscr{S}}_{\mu|S}^{IV}$ are associated with 16 different m-specifications. We thus obtain 96 different m-specifications, while the overall numbers of microscopic states is 128 because of assumption FM\ref{fm3}.

Let us consider now the \emph{detection probability} (or \emph{intrinsic efficiency}) of a measurement of a spin observable $\sigma_*(n)$ on a physical object $\alpha \in ext S$ according to $\mathscr{M}(3)$. To this end, let us observe that the existence of a unique d--distribution associated with each microscopic state $S_{\mu}$ implies that in $\mathscr{M}(3)$ one can calculate such a detection probability as the ratio $N^d(\sigma_*(n))/N$, where $N$ (=128) is the overall number of states and $N^d(\sigma_*(n))$ is the number of microscopic states such that $\alpha$ is detected in the measurement of $\sigma_*(n)$. Then we obtain by inspection that this probability is $1$ if $*=z$, $1/2$ if $*=x,y$. The detection probabilities for the measurements of a pair or a triple of compatible spin observables 
follow at once by noticing that the uniqueness of the d--distribution associated with $S_{\mu}$ implies that the detection mappings are independent random variables in $\mathscr{M}(3)$.\footnote{\label{foot}We note that these intrinsic efficiencies are very low. We however do not consider this feature of the model as a problem. Indeed, the intrinsic detection efficiencies are free parameters in the ESR model, whose upper limits are generally much higher \cite{gs10,gs12}. As we have anticipated in the Introduction, our models have the nontrivial aim of illustrating how the theory works and recovering some local finite models for the GHZ experiment that have been proposed in the literature, but do not intend to supply a realistic  description of what is actually going on in the GHZ experiment.} Moreover, it is easy to verify by inspection that $\mathscr{M}(3)$ satisfies condition DM (this condition has indeed been a guide for its construction).

Let us come to conditional on detection probabilities and let us show that they satisfy the adequacy condition AC. To this end, let us observe that the uniqueness of the d--distribution associated with $S_{\mu}$ implies that the conditional on detection probability of obtaining a prefixed set of outcomes in a set of compatible measurements on a physical object $\alpha \in ext S$ is given by the ratio $m/M$ where $M$ is the number of microscopic states such that $\alpha$ is detected in every measurement of the set and $m$ is the number of these microscopic states such that $\alpha$ yields the prefixed outcomes. Then, let us consider the measurement of an observable $\sigma_*(n)$ on $\alpha$. We obtain by inspection the following results. 

If $*=x,y$, for every $n$ one gets that there are 64 microscopic states in $\tilde{\mathscr{S}}_{\mu|S}$ such that $\lambda_{*n}(S_\mu)=D$. Moreover, 32 of them are such that $v(\sigma_*(n))=+1$ and 32 such that $v(\sigma_*(n))=-1$. Hence both conditional on detection probabilities of the $+1$ and $-1$ outcomes are $32/64=1/2$ and coincide with the probabilities predicted by QM.

If $*=z$, for every $n$ and $S_\mu\in\tilde{\mathscr{S}}_{\mu|S}$, Table 1 implies $\lambda_{zn}(\alpha)=D$. Hence, each particle in $\alpha$ is detected whenever a spin measurement along the $z-$axis is performed on it. Since the set $\tilde{\mathscr{S}}_{\mu|S}$ contains 64 microscopic states such that $v(\sigma_z(n))=+1$ and 64 microscopic states such that $v(\sigma_z(n))=-1$, both conditional on detection probabilities of the $+1$ and $-1$ outcomes are $1/2$ and coincide with the probabilities predicted by QM. 

Coming to joint measurements, one can see by inspection that the conditional on detection probability of obtaining a given combination of the outcomes $+1$ and $-1$ when measuring two arbitrary observables on two different particles of $\alpha$ 
has the value required by condition AC and specified in Sec.~\ref{family}.
Analogously, the conditional on detection probability of obtaining a given combination of the outcomes $+1$ and $-1$ when measuring three arbitrary observables on the three particles of $\alpha$ 
 whenever the triple does not coincide with one of the triples that appear in the measurements $M^{I}$, $M^{II}$, $M^{III}$ and $M^{IV}$ has the value required by condition AC. All these conditional on detection probabilities therefore coincide with the probabilities predicted by QM. Furthermore, let us consider the measurement $M^I$. One sees by inspection that there are $16$ microscopic states in $\tilde{\mathscr{S}}_{\mu|S}$ such that the three particles in $\alpha$ are detected (to be precise the states in $\tilde{\mathscr{S}}^{I0}_{\mu|S}$) and that for each of these states the obtained outcomes satisfy Eq.~(\ref{vI}). Hence the conditional on detection probability that this equation be satisfied is $16/16=1$, which coincides with the probability predicted by QM, consistently with condition AC. Furthermore, each triple of outcomes satisfying Eq.~(\ref{vI}) occurs in $4$ of the $16$ aforesaid states. Hence, it has conditional on detection probability $4/16=1/4$, which again coincides with the probability predicted by QM (see Eq.~(\ref{I})).
 Similar arguments apply when considering the measurements $M^{II}$, $M^{III}$ and $M^{IV}$.
 
To complete our task it remains to show that $\mathscr{M}(3)$ can be converted into one of the toy models constructed by Szab\'{o} and Fine. To this end, let us associate every m--specification $m(\alpha)$ with a \emph{combination} $c(\alpha)$, omitting the values of the spin components of the three particles along the z-axis and substituting the letter $D$ (which according to Szab\'{o} and Fine stands for ``defectiveness'') to $0$. Therefore the 16 microscopic states in $\tilde{\mathscr{S}}^{I0}_{\mu|S}$ are associated with 4 different combinations (e.g. $(+1,D,D,+1,D,+1)$, $(+1,D,D,-1,D,-1)$, etc.), and similarly the 16 microscopic states in $\tilde{\mathscr{S}}^{II0}_{\mu|S}$, or $\tilde{\mathscr{S}}^{III0}_{\mu|S}$, or $\tilde{\mathscr{S}}^{IV0}_{\mu|S}$ are associated with 4 different combinations. Furthermore, the 16 microscopic states in $\tilde{\mathscr{S}}^{I}_{\mu|S}\cap\tilde{\mathscr{S}}^{II}_{\mu|S}\cap\tilde{\mathscr{S}}^{III}_{\mu|S}$ are associated with 8 different combinations (e.g., $(D,+1,D,+1,D,+1)$, $(D,+1,D,+1,D,-1)$, etc.), and similarly the 16 microscopic states in $\tilde{\mathscr{S}}_{\mu|S}^{I} \cap \tilde{\mathscr{S}}_{\mu|S}^{II} \cap \tilde{\mathscr{S}}_{\mu|S}^{IV}$, or $\tilde{\mathscr{S}}_{\mu|S}^{I} \cap \tilde{\mathscr{S}}_{\mu|S}^{III} \cap \tilde{\mathscr{S}}_{\mu|S}^{IV}$, or $\tilde{\mathscr{S}}_{\mu|S}^{II} \cap \tilde{\mathscr{S}}_{\mu|S}^{III} \cap \tilde{\mathscr{S}}_{\mu|S}^{IV}$ are associated with 8 different combinations. We thus obtain the 48 combinations that, according to Szab\'{o} and Fine ``produce a triple detection coincidence at only one ... triad(s) of angles'', hence form a prism model for the GHZ experiment \cite{sf02}. 

Let us add some comments on probabilities. We recall that the set of all combinations constitutes the space $\Lambda$ of hidden variables on which a probability measure is defined in the Szab\'{o} and Fine toy model. By restricting $\Lambda$ to the set of combinations obtained above, we notice that the probability distribution induced on $\Lambda$ by $\mathscr{M}(3)$ is not uniform. Indeed, each combination corresponding to an m-specification in $\tilde{\mathscr{S}}_{\mu|S}^{I0}$, or $\tilde{\mathscr{S}}_{\mu|S}^{II0}$, or $\tilde{\mathscr{S}}_{\mu|S}^{III0}$, or $\tilde{\mathscr{S}}_{\mu|S}^{IV0}$ can be obtained from 4 microscopic states: hence, the probability that a physical object $\alpha$ in the macroscopic state $S$ is associated with such a combination is $4/128=1/32$. Instead each combination corresponding to an m-specification in one of the remaining sets of the partition in P3 can be obtained from 2 microscopic states only: hence the probability that $\alpha$ is associated with such a combination is $2/128=1/64$.

\subsection{$\mathscr{M}(1)$: a finite model with one detection failure}\label{3toy}
The features of the model $\mathscr{M}(3)$ lead one to wonder whether the family in Sec. \ref{family} contains finite local models satisfying the requirement that for every physical object only one detection mapping takes value $U$. It is then easy to see that a model of this kind cannot exist. Indeed, if $S_\mu$ belongs to one of the sets  $\tilde{\mathscr{S}}_{\mu|S}^{I0}$, $\tilde{\mathscr{S}}_{\mu|S}^{II0}$, $\tilde{\mathscr{S}}_{\mu|S}^{III0}$ and $\tilde{\mathscr{S}}_{\mu|S}^{IV0}$, then for every $\alpha \in ext S_\mu$ more than one detection mapping must take value $U$ to avoid contradiction with the predictions of QM. Nevertheless, if one restricts the requirement above to the physical objects in microscopic states that belong to $\tilde{\mathscr{S}}_{\mu|S}\setminus(\tilde{\mathscr{S}}_{\mu|S}^{I0} \cup \tilde{\mathscr{S}}_{\mu|S}^{II0} \cup \tilde{\mathscr{S}}_{\mu|S}^{III0} \cup \tilde{\mathscr{S}}_{\mu|S}^{IV0})$, then finite local models for the GHZ experiment can be constructed. For instance, one can construct a model, that we agree to call $\mathscr{M}(1)$, by assuming that, for every $S_\mu\in\tilde{\mathscr{S}}_{\mu|S}^{I0} \cup \tilde{\mathscr{S}}_{\mu|S}^{II0} \cup \tilde{\mathscr{S}}_{\mu|S}^{III0} \cup \tilde{\mathscr{S}}_{\mu|S}^{IV0}$ and $\alpha \in ext S_{\mu}$, $\lambda(\alpha)= (U,U,U;\,U,U,U;\,U,U,U)$ (hence $m(\alpha)=(0,0,0;\,0,0,0;\,0,0,0)$, which implies that $\alpha$ is never detected if a measurement of $\sigma_*(n)$ 
is performed on it), and adding Table~2 to complete the model.\footnote{Because of lack of space we do not report in Table~2 the outcomes in the measurements $M^I$, $M^{II}$, $M^{III}$ and $M^{IV}$; these can be easily deduced, however from the m--specifications that are listed in Table~2.} We then see by inspection that for every $S_\mu \in  \tilde{\mathscr{S}}_{\mu|S}^{I} \cap \tilde{\mathscr{S}}_{\mu|S}^{II} \cap \tilde{\mathscr{S}}_{\mu|S}^{III}$ there are three different d--distributions, each leading to a different m-specification, and that different states in $\tilde{\mathscr{S}}_{\mu|S}^{I} \cap \tilde{\mathscr{S}}_{\mu|S}^{II} \cap \tilde{\mathscr{S}}_{\mu|S}^{III}$ lead to different m-specifications. Hence the 16 microscopic states in $\tilde{\mathscr{S}}_{\mu|S}^{I} \cap \tilde{\mathscr{S}}_{\mu|S}^{II} \cap \tilde{\mathscr{S}}_{\mu|S}^{III}$ lead to 48 different m-specifications. Similarly, the 16 microscopic states in each of the sets $\tilde{\mathscr{S}}_{\mu|S}^{I} \cap \tilde{\mathscr{S}}_{\mu|S}^{II} \cap \tilde{\mathscr{S}}_{\mu|S}^{IV}$, $\tilde{\mathscr{S}}_{\mu|S}^{I} \cap \tilde{\mathscr{S}}_{\mu|S}^{III} \cap \tilde{\mathscr{S}}_{\mu|S}^{IV}$ and $\tilde{\mathscr{S}}_{\mu|S}^{II} \cap \tilde{\mathscr{S}}_{\mu|S}^{III} \cap \tilde{\mathscr{S}}_{\mu|S}^{IV}$ lead to 48 different m-specifications. Every d--distribution and m-specification, however, occurs twice in Table~2. Hence, Table~2 contains 6 different d--distributions and 96 different m-specifications, while the overall number of microscopic states is 64.

\begin{figure} 
\begin{center}
\includegraphics[width=0.6\textwidth]{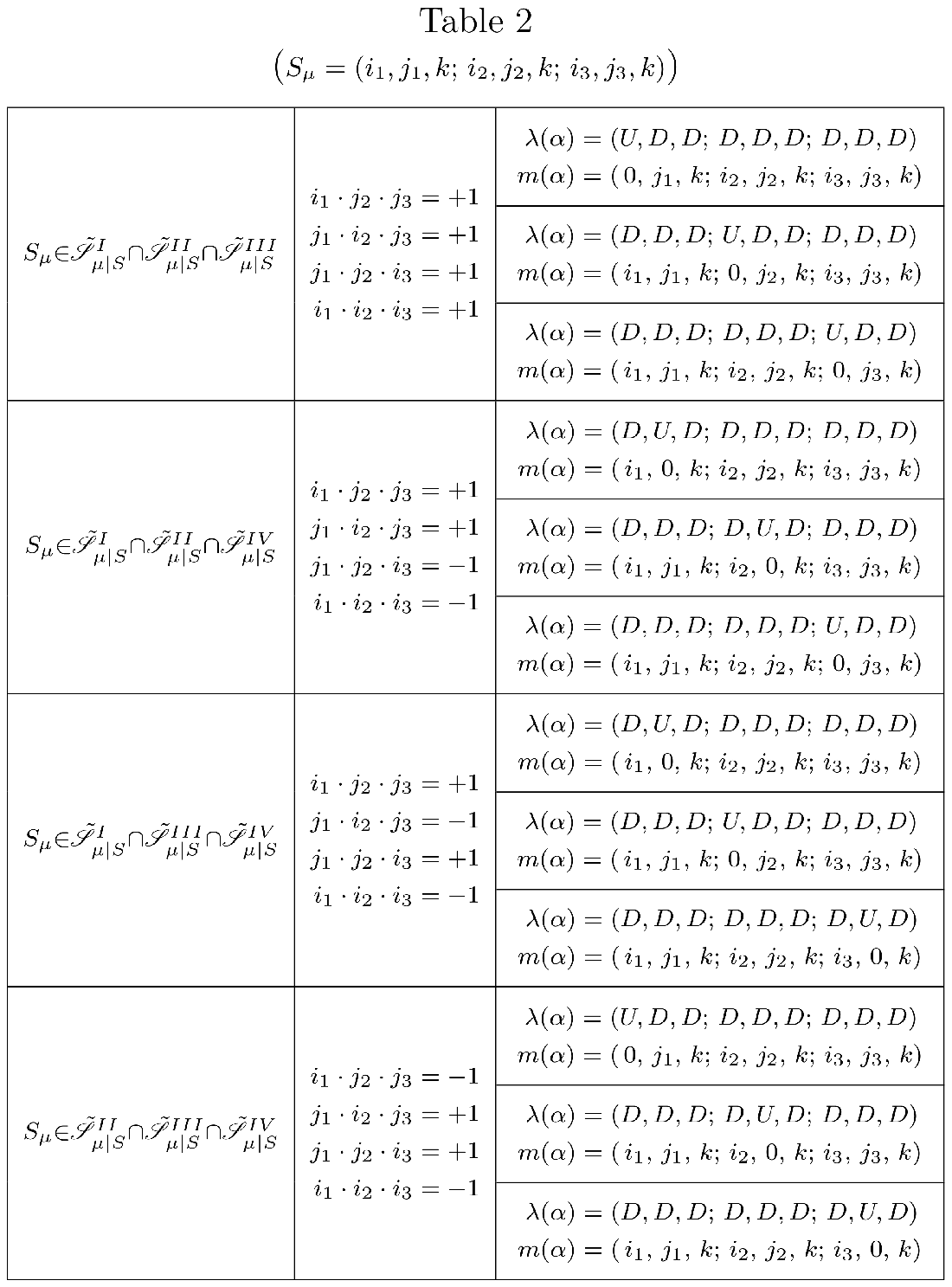}
\end{center}
\end{figure}

Let us discuss now the detection probability of a set of compatible measurements of spin observables on a physical object $\alpha \in ext S$ according to $\mathscr{M}(1)$. To this end, let us observe that the random variables $\lambda_{x1}$, $\lambda_{y1}$, $\lambda_{x2}$, $\lambda_{y2}$, $\lambda_{x3}$ and $\lambda_{y3}$ are not independent in $\mathscr{M}(1)$ (for instance, if $\alpha \in ext S_\mu$, with $S_{\mu}\in\tilde{\mathscr{S}}_{\mu|S}^{I} \cap \tilde{\mathscr{S}}_{\mu|S}^{II} \cap \tilde{\mathscr{S}}_{\mu|S}^{III}$ and $\lambda_{x1}(\alpha)=U$, then $\lambda_{x2}(\alpha)=\lambda_{x3}(\alpha)=D$). Assumption FM\ref{fm4}, (ii), then implies that the probability of $\lambda_{*n}(\alpha)=D$ whenever some prefixed detection mappings different from $\lambda_{*n}$ take value $D$ on $\alpha$ is given by the ratio $M^\alpha/M$, where $M$ is the number of d--distributions that assign value $D$ on $\alpha$ to the prefixed detection mappings and $M^\alpha$ is the number of these d--distributions that are also such that $\lambda_{*n}(\alpha)=D$. Hence we can calculate the detection probabilities for any set of compatible measurements on a physical object $\alpha$ in the microscopic state $S_\mu$ and show, in particular, that the model in Table~2 satisfies condition DM.\footnote{Let us provide some instances. Let $S_\mu\in\tilde{\mathscr{S}}_{\mu|S}^{I} \cap \tilde{\mathscr{S}}_{\mu|S}^{II} \cap \tilde{\mathscr{S}}_{\mu|S}^{III}$. If a measurement of $\sigma_x(1)$ is performed, the detection probability is $2/3$; if a measurement of $\sigma_y(1)$ is performed it is $1$. If measurements of $\sigma_x(1)$ and $\sigma_y(2)$ are performed, the detection probability is $2/3\cdot 1\!\!=\!2/3$; if measurements of $\sigma_x(1)$ and $\sigma_x(2)$ are performed it is $2/3\cdot 1/2=1/3$. If measurements of $\sigma_x(1)$, $\sigma_x(2)$ and $\sigma_x(3)$ are performed, the detection probability is $2/3\cdot 1/2 \cdot 0=0$; if measurements of $\sigma_x(1)$, $\sigma_y(2)$ and $\sigma_x(3)$ are performed it is $2/3\cdot 1\cdot 1/2=1/3$; if measurements of $\sigma_x(1)$, $\sigma_y(2)$ and $\sigma_y(3)$ are performed, it is $2/3\cdot 1\cdot 1=2/3$.}
Because of assumption FM\ref{fm4}, (i), we can then calculate the overall detection probability for any set of compatible measurements on a physical object $\alpha$ in the state $S$ as the ratio between the sum of all these detection probabilities (one for each microscopic state $S_\mu\in\tilde{\mathscr{S}}_{\mu|S}$) and the number N of microscopic states.\footnote{For instance, if a measurement of $\sigma_*(n)$ is performed, the detection probability is $1/128\cdot(2/3\cdot 16+2/3\cdot 16+16+16)=5/12$ if $*=x,y$, it is $1/2$ if $*=z$, independently of $n$. Note that these intrinsic efficiencies are rather small, but we do not consider this feature of the model as a problem because of the same arguments advanced in footnote + with reference to $\mathscr{M}(3)$.} 

\begin{figure} 
\begin{center}
\includegraphics[width=0.8\textwidth]{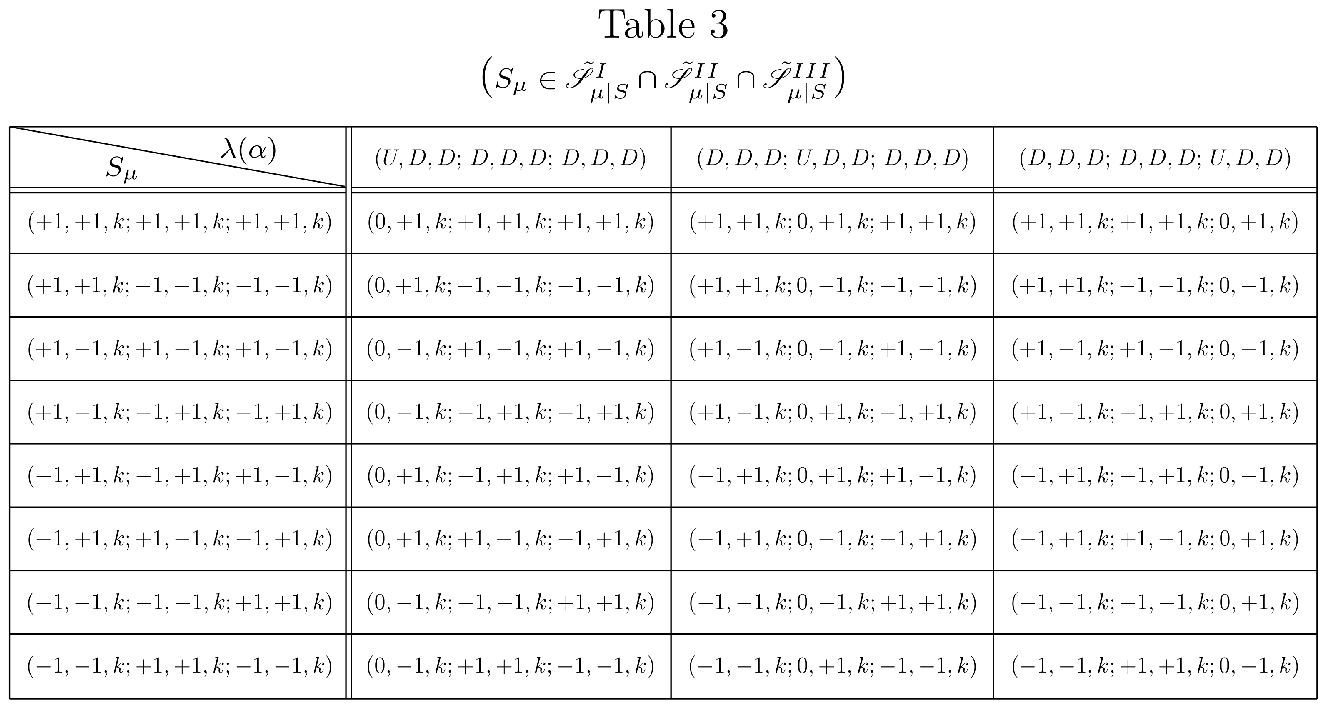}
\end{center}
\end{figure}

Let us come to the conditional on detection probabilities and let us show that they satisfy condition AC. To get this result it is expedient to refer to Table 3, which exhibits all the m--specifications that actually occur in the specific case $S_\mu\in\tilde{\mathscr{S}}_{\mu|S}^{I} \cap \tilde{\mathscr{S}}_{\mu|S}^{II} \cap \tilde{\mathscr{S}}_{\mu|S}^{III}$, bearing in mind that similar tables can be drawn when  $S_{\mu} \in\tilde{\mathscr{S}}_{\mu|S}^{I} \cap \tilde{\mathscr{S}}_{\mu|S}^{II} \cap \tilde{\mathscr{S}}_{\mu|S}^{IV}$, $S_{\mu} \in \tilde{\mathscr{S}}_{\mu|S}^{I} \cap \tilde{\mathscr{S}}_{\mu|S}^{III} \cap \tilde{\mathscr{S}}_{\mu|S}^{IV}$ and $S_{\mu} \in \tilde{\mathscr{S}}_{\mu|S}^{II} \cap \tilde{\mathscr{S}}_{\mu|S}^{III} \cap \tilde{\mathscr{S}}_{\mu|S}^{IV}$. Indeed, statements (i) and (ii) in assumption FM\ref{fm4} now imply that the probability that a physical object $\alpha$ in the state $S$ is associated by $\lambda$ with a given m--specification listed in one of these tables is $64/128$ (the probability that the microscopic state $S_{\mu}$ of $\alpha$ belongs to $\tilde{\mathscr{S}}_{\mu|S} \setminus (\tilde{\mathscr{S}}_{\mu|S}^{I0} \cup \tilde{\mathscr{S}}_{\mu|S}^{II0} \cup \tilde{\mathscr{S}}_{\mu|S}^{III0} \cup \tilde{\mathscr{S}}_{\mu|S}^{IV0})$) times 2/192 (the probability of the given m--specification whenever $S_{\mu}$ belongs to the foregoing set), that is, it is $1/192$. Hence the conditional on detection probability of obtaining a prefixed set of nonzero outcomes in a set of compatible measurements of $\alpha$ is given by the ratio $m/M$, where $M$ is the number of m--specifications where a nonzero outcome occurs for each measurement in the set and $m$ is the number of these m--specifications where the prefixed outcomes occur.

Let us consider now the measurement of an observable $\sigma_*(n)$ on $\alpha$. We obtain by inspection the following results.

If $*=x,y$, for every $n$ there are $64$ m--specifications such that the outcome of the measurement is nonzero, with $32$ m--specifications such that $v(\sigma_*(n))=+1$ and $32$ m--specifications such that $v(\sigma_*(n))=-1$. Hence both conditional on detection probabilities of the $+1$ and $-1$ outcomes are $1/2$ and coincide with the probabilities predicted by QM. 

If $*=z$ there are $96$ m--specifications such that the outcome of measurement is nonzero, with $48$ m--specifications such that $v(\sigma_z(n))=+1$ and $48$ m--specifications such that $v(\sigma_z(n))=-1$. Hence both conditional on detection probabilities of the $+1$ and $-1$ outcomes are $1/2$ and coincide with the probabilities predicted by QM.

Coming to joint measurements, one can see by inspection that the conditional on detection probability of obtaining a given combination of the outcomes $+1$ and $-1$ when measuring two arbitrary observables on two different particles of $\alpha$ 
has the value required by condition AC and specified in Sec.~\ref{family}.
Analogously, the conditional on detection probability of obtaining a given combination of the outcomes $+1$ and $-1$ when measuring three arbitrary observables on the three particles of $\alpha$ 
whenever the triple does not coincide with one of the triples that appear in the measurements $M^{I}$, $M^{II}$, $M^{III}$ and $M^{IV}$ has the value required by condition AC. All these conditional on detection probabilities therefore coincide with the probabilities predicted by QM. Furthermore, let us consider the measurement $M^I$. One sees by inspection that there are $48$ m--specifications such that the three particles in $\alpha$ are detected, and that for each of these m--specifications the obtained outcomes satisfy Eq.~(\ref{vI}). Hence the conditional on detection probability that this equation be satisfied is $48/48=1$, which coincides with the probability predicted by QM, consistently with condition AC. Furthermore, each triple of outcomes satisfying Eq.~(\ref{vI}) occurs in $12$ of the $48$ aforesaid m--specifications. Hence, it has conditional on detection probability $12/48=1/4$, which again coincides with the probability predicted by QM (see Eq.~(\ref{I})). Similar arguments apply when considering the measurements $M^{II}$, $M^{III}$ and $M^{IV}$.

To complete our task it remains to show that $\mathscr{M}(1)$ can be converted into one of the toy models constructed by Szab\'{o} and Fine. To this end, let us proceed as in Sec.~\ref{1toy}, associating every m--specification $m(\alpha)$ with a \emph{combination} $c(\alpha)$ in which the values of the spin components of the three particles along the z-axis are omitted and the letter $D$ is substituted to $0$. Therefore the $48$ m--specifications in Table~3 are associated with $24$ different combinations (e.g. $(D,+1,+1,+1,+1,+1)$, $(+1,+1,D,+1,+1,+1)$, etc.). Similarly, one obtains $24$ different combinations from each table that can be drawn by referring to $\tilde{\mathscr{S}}_{\mu|S}^{I} \cap \tilde{\mathscr{S}}_{\mu|S}^{II} \cap \tilde{\mathscr{S}}_{\mu|S}^{IV}$, $\tilde{\mathscr{S}}_{\mu|S}^{I} \cap \tilde{\mathscr{S}}_{\mu|S}^{III} \cap \tilde{\mathscr{S}}_{\mu|S}^{IV}$ and $\tilde{\mathscr{S}}_{\mu|S}^{II} \cap \tilde{\mathscr{S}}_{\mu|S}^{III} \cap \tilde{\mathscr{S}}_{\mu|S}^{IV}$. These $96$ combinations appear twice in the Tables, hence we are left with the $48$ combinations that occur in Table 1 of the Szab\'{o} and Fine paper \cite{sf02}.

Let us add some comments on probabilities. We recall that the set of all combinations constitutes the space $\Lambda$ of hidden variables on which a probability measure is defined in the Szab\'{o} and Fine toy model. By restricting $\Lambda$ to the set of combinations obtained above, we notice that the probability distribution induced on $\Lambda$ is uniform in $\mathscr{M}(1)$, consistently with the assumption introduced by Szab\'{o} and Fine to obtain a model with maximal triple detection efficiency.

\subsection{$\mathscr{M}$(2): a finite model with two detection failures}\label{2toy}
\begin{figure} 
\begin{center}
\includegraphics[width=0.6\textwidth]{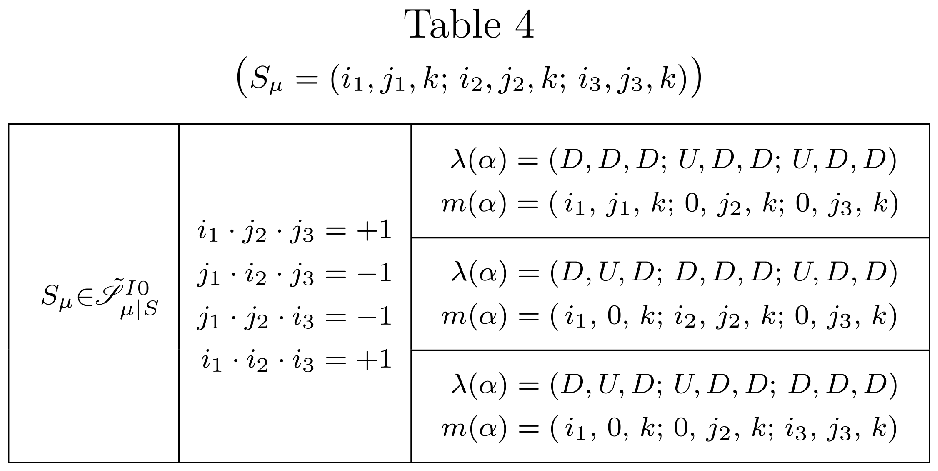}
\end{center}
\end{figure}

\begin{figure} 
\begin{center}
\includegraphics[width=0.6\textwidth]{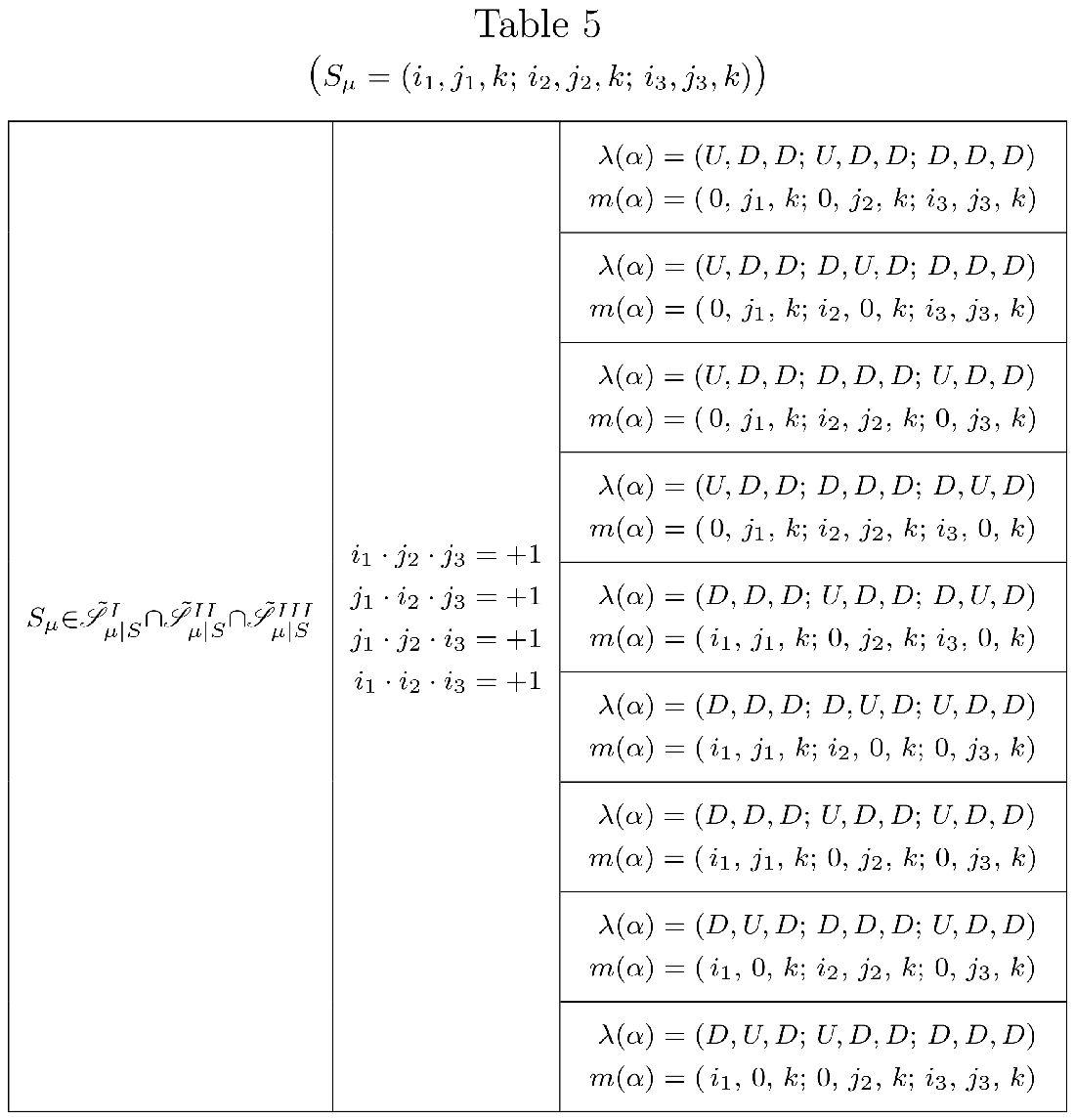}
\end{center}
\end{figure}

Finally,
Table 4 and Table 5 illustrate in the cases $S_\mu\in\tilde{\mathscr{S}}_{\mu|S}^{I0}$
and $S_\mu\in\tilde{\mathscr{S}}_{\mu|S}^{I} \cap \tilde{\mathscr{S}}_{\mu|S}^{II} \cap \tilde{\mathscr{S}}_{\mu|S}^{III}$, respectively, a finite model for the GHZ experiment, that we agree to call $\mathscr{M}(2)$, in which there are different d--distributions associated with each microscopic state $S_\mu$ belonging to a given element of the partition in P\ref{p3}, as in $\mathscr{M}(1)$, but there are two detection mappings that take value $U$. Of course, tables similar to Table 4 hold when considering $\tilde{\mathscr{S}}_{\mu|S}^{II0}$, $\tilde{\mathscr{S}}_{\mu|S}^{III0}$ and $\tilde{\mathscr{S}}_{\mu|S}^{IV0}$ in place of $\tilde{\mathscr{S}}_{\mu|S}^{I0}$, and tables similar to Table 5 hold when considering $\tilde{\mathscr{S}}_{\mu|S}^{I} \cap \tilde{\mathscr{S}}_{\mu|S}^{II} \cap \tilde{\mathscr{S}}_{\mu|S}^{IV}$, $\tilde{\mathscr{S}}_{\mu|S}^{I} \cap \tilde{\mathscr{S}}_{\mu|S}^{III} \cap \tilde{\mathscr{S}}_{\mu|S}^{IV}$ and $\tilde{\mathscr{S}}_{\mu|S}^{I} \cap \tilde{\mathscr{S}}_{\mu|S}^{III} \cap \tilde{\mathscr{S}}_{\mu|S}^{IV}$ in place of $\tilde{\mathscr{S}}_{\mu|S}^{I} \cap \tilde{\mathscr{S}}_{\mu|S}^{II} \cap \tilde{\mathscr{S}}_{\mu|S}^{III}$. The m-specifications that occur in the first series of tables are all different, and each table contains 48 m--specifications: hence there are in these tables 192 m-specifications. The m-specifications that occur in the second series of tables coincide with the specifications that occur in the first series (compare in particular the last three m-specifications in Table 5 with the m-specifications in Table 4), and every m-specification occurs 4 times if all tables are considered. Therefore the overall number of different m-specifications is 192. By omitting the values of the spin components of the three particles along the z--axis and substituting the letter $D$ to $0$, as we did in Secs. \ref{1toy} and \ref{3toy}, we obtain the 96 combinations that, according to Szab\'o and Fine, produce a triple detection coincidence at two triads of angles, forming a prism model for the GHZ experiment \cite{sf02}. For the sake of brevity we avoid to calculate explicitly in this case the detection probabilities and to show that the conditional on detection probabilities satisfy condition AC. This result can be easily obtained, however, by direct inspection, following the paradigms provided in Secs. \ref{1toy} and \ref{3toy}.

\end{document}